\documentstyle[12pt]{article}
\renewcommand{\beta}{\sigma}
\newcommand{\ber}{\begin{eqnarray}}
\newcommand{\eer}{\end{eqnarray}}
\newcommand{\bea}{\begin{equation}}
\newcommand{\eea}{\end{equation}}
\begin{document}
\title{\bf The Non-local Kardar-Parisi-Zhang Equation With Spatially Correlated Noise}
\author{Amit Kr. Chattopadhyay  \\
Department of Theoretical Physics \\
Indian Association For the Cultivation of Science \\
Jadavpur, Calcutta 700 032, India \\
e-mail: tpac2@mahendra.iacs.res.in}
\date{}
\maketitle
\begin{abstract}
The effects of spatially correlated noise on a phenomenological
equation equivalent to a non-local version of the Kardar-Parisi-Zhang
(KPZ) equation are studied via the dynamic renormalization
group (DRG) techniques. 
The correlated noise coupled with the long ranged nature of
interactions prove the existence of different phases in different
regimes, giving rise to a range of roughness exponents defined by
their corresponding critical dimensions. Finally self-consistent
mode analysis is employed to compare the non-KPZ
exponents obtained as a result of the long range-long range
interactions with the DRG results.\\\\ 

05.40.+j,05.70.Ln,64.60.Ht,68.35.Fx\\

\end{abstract}
\newpage 
The interests in the non-equilibrium growth mechanisms in the formation of surfaces and interfaces as also the description of directed polymers, bacterial growth, etc. and of late the protein folding problems, although apparently being the representation of different physical processes, have all been encapsulated in one single non-linear continuum equation, the much celebrated Kardar-Parisi Zhang equation [1]. The notion of universality classes defined by this standard KPZ equation [1] suggests the existence of a phase transition from the EW class [2] to the non-linear KPZ class above a particular critical dimension ($ d>2 $). However although this stochastic equation has by now become a model of dynamic critical phenomena for a vast range of growth problems, the fact remains that the basic non-linearity studied here is of a local nature, that is to say the growth occurs along a continuously varying local normal. 
\par
To incorporate the long-ranged nature of interactions which are necessary for a wide class of problems, eg. the long-ranged hydrodynamic interactions [12-14], proteins, colloids [15-18], etc. SM and SMB [3] developed a Langevin type equation studying the effects of the long-ranged (LR) feature of an evolving surface going beyond the local description of the KPZ non-linearity for the case of white noise. There the approach essentially consisted in introducing a term in the basic Langevin equation capable of correlating each site of the growing surface with all other sites. The objective was the transformation of the local non-linear term representing the lateral growth beyond the strict local description such that the correlation length now becomes at least of the system size. Still there remains to be seen the effects of interaction of correlated colored noise with the KPZ or KPZ-type non-linearity. 
\par
The results of non-white noise for the growth of rough interface has been generalized by Medina et al [4]. In two remarkable papers Chekhlov and Yakhot [5,6] and Hayot and Jayaprakash [7] have observed the effects of correlated noise for the one-dimensional Burgers' equation. They explored the occurence of shocks as well as the large distance, long time statistics of the fluctuations. Working along this line Frey, Tauber and Janssen [8] have also resorted to the long-ranged description of noise in their treatment of the scaling regimes and critical dimensions in the standard KPZ problem as well as in the conserved case [9]. Actually in all these cases, the nature of the noise is determined from the fact that to maintain turbulence in the flow, energy has to be supplied at large length scales near the boundaries and the consequent Kolmogorov type of dimensional argument brings about a spatial dependence in the noise correlation.
\par
Starting with the non-local equation proposed in [3], we have gone one step further in putting forth the effects of a non-white spatially correlated noise akin to that used in [4], the objective being the analysis of special features of interaction of long-ranged nature of KPZ type non-linearity with the long-ranged spatial correlation in noise. Use is made of dynamic renormalization group (DRG) techniques in arriving at the dynamic exponents, etc. We see that even for $ d<d_c $, both weak and strong noise interacting both with the local and non-local nature of the non-linearity give a range of critical exponents spanning a four dimensional space in terms of the dimensionless interaction strengths. Finally we reconfirm our DRG results from the self-consistent mode power counting arguments in the line developed in [10,11]. 
\par
The starting point of our analysis is the equation:

\bea 
\frac{\partial h(\vec r,t)}{\partial t}\:=\:\nu\:\nabla^2 h(\vec
r,t)\:+\:\frac{1}{2}\:\int\:d\vec r^{\prime}\:v(\vec r^{\prime})\:\vec 
\nabla h(\vec r+\vec r^{\prime},t).\vec \nabla h(\vec r-\vec
r^{\prime},t)\:+\:\eta(\vec r,t) 
\eea

where $ \nu $ is the diffusion constant
and $ \eta(\vec r,t) $ is the noise defined by

\bea
<\:\eta(\vec k,\omega) \eta(\vec
k^{\prime},\omega^{\prime})\:>\:=\:2D(\vec k)\:\delta^d (\vec k+\vec
k^{\prime})\:\delta(\omega+\omega^{\prime}) 
\eea

Again going by the prescription of [3], the kernel $ v(\vec r) $ 
has a short-ranged part $ \lambda_0 \delta^d (\vec r) $ and a
long-ranged part ~ $ r^{\rho-d} $. In Fourier space, 

\bea
v(\vec k)\:=\:\lambda_0\:+\:\lambda_{\rho} k^{- \rho}
\eea

All standard KPZ results are expected for $ \lambda_{\rho}=0 $. Our
introduction of the long-ranged (LR) version of the noise in coupling with
the long-ranged (LR) part in the interaction spectrum is expected to define
a legion of continuously varying exponents for a set of new non-KPZ
fixed points.
\par
The two exponents of interest, the roughness exponent $ \alpha $ and
the dynamic exponent z come along with the two-point height
correlation function in the hydrodynamic limit $ (\vec k,\omega)
\rightarrow 0 $,

\bea 
<\:h(\vec k,\omega) h(\vec k^{\prime},\omega^{\prime})\:>\:\sim\:
{\mid \vec k \mid}^{z-d-2\alpha}\:\delta^d (\vec k+\vec
k^{\prime})\:\delta(\omega+\omega^{\prime})\:f(\frac{\omega} {{\mid
    \vec k \mid}^z})
 \eea
 
 All information regarding the dynamic universality class of the phase
 will be contained in this $ \alpha $ and z.
\par
At d=1, the values of $ \alpha(=\frac{1}{2}) $ and $ z(=\frac{3}{2}) $
can be exactly determined. But at d=2, there is a transition from the
Gaussian fixed point (EW) and the non-linearity grows under rescaling.
Simple scaling from $ \vec x \rightarrow b \vec x, h \rightarrow
b^{\alpha} h $ and $ t \rightarrow b^z t $ shows that both the short
(SR) and long-ranged (LR) contributions in the interaction kernel are relevant for $ d<2 $ (where by SR interaction we mean the standard KPZ type non-linearity, that is the kernel is a constant multiplying $ \delta^d(\vec r) $ and the LR interaction implies a non-KPZ type $ \vec r $ dependent part) . Under this scale transformation the parameters of the
equation change by $ \nu \rightarrow b^{z-2} \nu, \lambda_0
\rightarrow b^{\alpha+z-2} \lambda_0, \lambda_{\rho} \rightarrow
b^{\alpha+z+\rho-2} \lambda_{\rho} $. If the noise strength $ D(\vec
k) $ in the hydrodynamic limit ($ \vec k,\omega \rightarrow 0 $)
is given by 
$ D(\vec k)\:=\:D_0\:+\:D_{\beta} k^{-2\beta} $, then $ D_0 \rightarrow
b^{-d-2\alpha+z} D_0 $ and $ D_{\beta} \rightarrow
b^{-4\beta-d-2\alpha+z} D_{\beta} $. For $
2\:<d\:<\:d_c=2+2\rho-4\beta, \lambda_{\rho} $ is relevant
at the EW fixed point (z=2) and for $ \rho>0 $, a non-KPZ fixed spectrum should be the outcome. The following DRG analysis gives a horizon of
unfounded results which shrink to the known results in d=1 and 2 as in
[3], new stable points are obtained and furthermore the introduction of the non-local noise provides even more complexity in the
interaction.
\par
The generalized Green's function 
$ G(\vec k,\omega)=\frac{<\frac{\partial h(\vec k,\omega)}{\partial
\eta(\vec k^{\prime},\omega^{\prime})}>}{{\delta}^d (\vec k+\vec k^{\prime})\:\delta(\omega+{\omega}^{\prime})} $ in the symmetrized Fourier space is given by,

\ber
G(\vec k,\omega)\:&=&\:G_0(\vec k,\omega)\:+\:4.(\frac{1}{2})^2\:{G_0}^2
(\vec k,\omega)\:\int\:\frac{d^d k^{\prime}}{{2\pi}^d}\:\frac{d\omega^{\prime}}
{2\pi}\:\vec k^{\prime}_{+}.\vec k^{\prime}_{-} \nonumber \\
& & v(2\vec k^{\prime})\:v(\frac{3\vec k}{2}+\vec k^{\prime}) 
G_0(\vec k^{\prime}_{-},\omega_{-}^{\prime})\:G_0(\vec k^{\prime}_{+},\omega^{\prime})\nonumber\\
& &G_0(-\vec k^{\prime}_{+},-\omega_{+}^{\prime})\:2D(\vec k^{\prime}_{+},\omega_{+}^{\prime})
\eer

where symbolically $ X_{\pm} = \frac{X}{2}\:\pm\:X^{\prime} $, with $
X=\vec p $ or $ \omega $.  $ G_0(\vec k,\omega) = -i \omega + \nu k^2
$ represents the unperturbed term in the propagator expansion. We
consider diagrams upto the one-loop order $ O(v^2) $ and follow this
expansion with a detailed dynamic renormalization procedure as in
Medina, et al [4]. The propagator expansion gives us the viscosity
flow, the vertex renormalization gives the flows of $ \lambda_0 $ and
$ \lambda_{\rho} $ and the noise renormalization gives the $ D_0 $ and
$ D_{\beta} $ flows. As usual we integrate out the fast modes in the
momentum shell $ \Lambda e^{-l}\:\leq\:k\:\leq\:\Lambda $, set the
cut-off $ \Lambda=1 $ and develop the flow equations of the rescaled
parameters around the fixed points.
\par
Due to the Galilean invariance of eqn.(1), $ \lambda_0 $
is not renormalized. The RG transformation being analytic in nature, $
\lambda_{\rho} $ is also not renormalized, only the Galilean identity
is modified in this case (2-$ \rho $ instead of 2),

\bea
\alpha+z\:=\:2-\rho
\eea

where $ \rho=0 $ for $ \lambda_0 $ flow. From the above
considerations, we get the following flow equations for $ \nu $, D's
and $ \lambda's $,

\bea
\frac{d\nu}{dl}\:=\:\nu\:[z-2-K_d\:\frac{D(1)v(2)v(1)}
{\nu^3}\:\frac{d-2+f(1)+3g(1)}{4d}]
\eea
\bea
\frac{dD(k)}{dl}\:=\:D(k)\:[z-2\alpha-d-f(k)]\:+\:K_d\:
\frac{D^2(1)}{4{\nu}^3}\:{v^2(2)}
\eea
\bea
\frac{d{{\lambda}_x}}{dl}\:=\:{\lambda_x}\:[\alpha+z-2+x]
\eea

where $ x=0 $ or $ \rho $ respectively.
Here $ f(q)=\frac{\partial ln D(k)}{\partial ln k}\mid_{k=q} $ and $ g(q)=
\frac{\partial ln v(k)}{\partial ln k}\mid_{k=q} $, $ K_d=\frac{S_d}
{(2\pi)^d} $, $ S_d $ representing the d-dimensional surface of a unit
d+1 dimensional sphere.
\par
In terms of the dimensionless interaction strengths
$ U^2_{0,\beta} = K_d \frac{{\lambda_0}^2 D_{\beta}}{\nu^3} $, where short-
ranged interaction couples with long-ranged noise (the Medina, et al zone)
and similar other parameters $ U^2_{\rho,\beta} = K_d \frac{{\lambda_
{\rho}}^2 D_{\beta}}{\nu^3} $,
$ U^2_{0,0} = K_d \frac{{\lambda_0}^2 D_0}{\nu^3} $ (ordinary KPZ case),
$ U^2_{\rho,0} = K_d \frac{{\lambda_{\rho}}^2 D_0}{\nu^3} $ (the SM and SMB
zone), the flow equations can be combined to give,

\ber
\frac{dU_{0,\beta}}{dl}\:&=&\:[\frac{2-d+2\beta}{2}]U_{0,\beta}\:+\:
3[\frac{d-2-2\beta}{8d}]U^3_{0,\beta}\:+\nonumber \\
& & \frac{U_{0,\beta}}{8d}\:[3(d-2)U^2_{0,0}\:+\:3.
2^{-\rho}(d-2-3\rho)U^2_{\rho,0}\:+\nonumber \\
& & \:3(1+2^{-\rho})(d-2)U_{0,0}U_{\rho,0}\:+\:3.
2^{-\rho}(d-2-2\beta-3\rho) U^2_{\rho,\beta}\nonumber \\
& &+\:3(1+2^{-\rho})(d-2-2\beta)U_{0,\beta}U_{\rho,\beta}]
\eer
\ber
\frac{dU_{\rho,\beta}}{dl}\:&=&\:[\frac{2-d+2\beta+2\rho}
{2}]U_{\rho,\beta}\:+\:3.2^{-\rho}[\frac{d-2-2\beta-3\rho}{8d}]U^3_
{\rho,\beta}\:+\nonumber \\
& &  \frac{U_{\rho,\beta}}{8d}\:[3(d-2)U^2_{0,0}\:+\:3.
2^{-\rho}(d-2-3\rho)U^2_{\rho,0}\:+\nonumber \\
& & \:3(1+2^{-\rho})(d-2)U_{0,0}U_{\rho,0}\:+\:3.
(d-2-2\beta)U^2_{0,\beta}\:+\nonumber\\
& &3(1+2^{-\rho})(d-2-2\beta)U_{0,\beta}U_{\rho,\beta}]
\eer
\ber
\frac{dU_{0,0}}{dl}\:&=&\:[\frac{2-d}{2}]U_{0,0}\:+\:[\frac
{2d-3}{4d}]U^3_{0,0}\:+\nonumber \\
& & \frac{U_{0,0}}{8d}\:[2^{-\rho}\{(3+2^{-\rho})d-6-
9\rho\}U^2_{\rho,0}\:\nonumber\\
& &+\:\{3(1+2^{-\rho})(d-2)+d.2^{-\rho+1}\}U_{0,0}U_{\rho,0}\:+
3(d-2-2\beta)U^2_{0,\beta}\:\nonumber \\
& &+\:3.2^{-\rho}(d-2-2\beta-3\rho)U^2_{\rho,\beta}\:+\:3(1+2^{-\rho})
(d-2-2\beta)U_{0,\beta}U_{\rho,\beta}]
\eer
\ber
\frac{dU_{\rho,0}}{dl}\:&=&\:[\frac{2-d+2\rho}{2}]U_{\rho,
0}\:+\:[\frac{(3+2^{-\rho})d-6-9\rho}{8d}]
U^3_{\rho,0}\:+\nonumber \\
& & \frac{U_{\rho,0}}{8d}\:[(4d-6)U^2_{0,0}\:+\:\{3(1+2^
{-\rho})(d-2)+d.2^{-\rho+1}\}U_{0,0}
U_{\rho,0}\:+\nonumber \\
& &  3(d-2-2\beta)U^2_{0,\beta}\:+\:3.2^{-\rho}(d-2-2\beta-
3\rho)U^2_{\rho,\beta}\:+\nonumber \\
& &3(1+2^{-\rho})(d-2-2\beta)U_{0,\beta}U_{\rho,\beta}]
\eer
Now let us define two sets of parameters: $
R_0=\frac{U_{0,0}}{U_{\rho,0}},\:
R_{\beta}=\frac{U_{0,\beta}}{U_{\rho,\beta}} $ and
$ S_0=\frac{U_{0,0}}{U_{0,\beta}},\:S_{\rho}
=\frac{U_{\rho,0}}{U_{\rho,\beta}} $. 
\par
We see that $ \frac{dR_y}{dl}=-\rho R_y $, where y=0, $ \beta $ and
consequently this rules out any off-axial fixed point in the $ R_0,
R_{\beta} $ parameter space (except for the trivial $ \beta=0 $ case).
\par 
In the $ U_{0,0},\:U_{\rho,0} $ plane, the axial fixed points are given by,

\bea
SR-SR  \equiv\:{{U^*}^2}_{\rho,0}=0,\:{{U^*}^2}_{0,0}=
\frac{2d(d-2)}{2d-3} ,\:\:\alpha+z=2 
\eea
 
\bea
LR-SR  \equiv\:{{U^*}^2}_{0,0}=0,\:{{U^*}^2}_{\rho,0}=
\frac{4(d-2-2\rho)}{2^{-\rho}\{(3+2^{-\rho})d-6-
9\rho\}}, \alpha+z=2-\rho. 
\eea

The first set (SR-SR) gives the well-known KPZ fixed point with 
$z=3/2,\:\alpha=1/2 $ for d=1. But the second set (LR-LR)gives the
non-KPZ behaviour and the results exactly match with [3] in this 
$U_{0,0},\:U_{\rho,0} $ plane:

\ber
z\mid_{{{U^*}^2}_{0,0}=0}\:&=&\:2\:+\:\frac{(d-2-2\rho)(d-2-3\rho)}{[(3+2^{-\rho})-6-9\rho]},\nonumber \\
\alpha\mid_{{{U^*}^2}_{0,0}=0}\:&=&\:-\rho-\frac{(d-2-2\rho)(d-2-3\rho)}{[(3+2^{-\rho})-6-9\rho]} 
\eer 

In diagram B of Fig.1, the dotted line gives an unstable zone
between $ d=\frac{9\rho+6}{3+2^{-\rho}} $ to $ d=2+2\rho $. Above the critical dimension, diagram C shows a smooth phase. For $ \rho=0 $,
all the LR fixed points go over to the SR ones.
\par
In the $ U_{0,\beta},\:U_{\rho,\beta} $ plane defined by $ R_{\beta}
$, there also are only two sets of axial fixed points:

\bea
SR-LR \equiv\:{{U^*}^2}_{\rho,\beta}=0,\:{{U^*}^2}_{0,\beta}=
\frac{4d}{3}
\eea

\bea
LR-LR \equiv\:{{U^*}^2}_{0,\beta}=0,
{{U^*}^2}_{\rho,\beta}=\frac{4d(2-d+2\beta+2\rho)}
{3.2^{-\rho}(2+2\beta+3\rho-d)}.
\eea

In this case the phase diagrams are exactly identical to Fig.1, only
the critical dimension is now modified to $ d_c=2+2\rho-4\beta $ and
the unstable zone now lies between $ d=2+2\beta+2\rho $ to $
d=2+2\beta+3\rho $ in this plane:

\ber
z\mid_{{{U^*}^2}_{0,\beta}=0}\:&=&\:\frac{1}{3}\:(4+d-2\beta-2\rho),\nonumber \\
{\alpha}\mid_{{{U^*}^2}_{0,\beta}=0}\:&=&\:-\rho+\frac{1}{3}\:(2-d+2\beta+2\rho).
\eer

\par
However a completely different qualitative behaviour is observed with
$ S_0 $ and $ S_{\beta} $. The $ S_0 $ flow equation reads $
\frac{dS_0}{dl}=S_0(-\beta+
\frac{1}{8}U^2_{0,0}+2^{-2\rho}U^2_{\rho,0}+\frac{2^{-\rho}}{4}
U_{0,0}U_{\rho,0} $ and as such off-axial fixed points exist in this
case for the 4-dimensional space of U's. But in the $
U_{0,0},\:U_{0,\beta} $ plane we get only axial fixed points where now
$ U_{0,0},\:U_{\rho ,0} $ =constants. Here the axial fixed points are

\bea
SR-SR \equiv\:{{U^*}^2}_{0,\beta}\:=\:0,\:{{U^*}^2}_
{0,0}\:=\:\frac{2d(d-2)}{2d-3}
\eea

\bea
SR-LR \equiv \:{{U^*}^2}_{0,0}\:=\:0,\:{{U^*}^2}_{0,
\beta}\:=\:\frac{4d}{3}
\eea

This gives 
\ber
z\mid_{{{U^*}^2}_{0,0}=0}\:&=&\:\frac{1}{3}\:(d+4-2\beta),\nonumber \\
{\alpha}\mid_{{{U^*}^2}_{0,0}=0}\:&=&\:\frac{1}{3}\:(2-d-
2\beta)
\eer

The results exactly match with the Medina, et al predictions and the phase diagram is given by FIG.2, X and Y are the axial and off-axial fixed points respectively. The point to be noted here is that unlike the previous two cases there is no unstable excluded region in the non-KPZ case.
\par
Although similar arguments as above apply for the $ S_{\rho} $ flow,
the most spectacular results are seen in the $
U_{\rho,0},\:U_{\rho,\beta} $ plane where the fixed points are given
by,

\bea
LR-SR \equiv\:{{U^*}^2}_{\rho,\beta}\:=\:0,\:{{U^*}^2}_
{\rho,0}\:=\:\frac{4d(2-d+2\rho)}{2^{-\rho}[6+9\rho-
(3+2^{-\rho})]}
\eea

\bea
LR-LR \equiv\:{{U^*}^2}_{\rho,0}\:=\:0,\:{{U^*}^2}_
{\rho,\beta}\:=\:\frac{4d(2-d+2\beta+2\rho)}{3.2^{-\rho}
(2+2\beta+3\rho-d)}.
\eea

Here for both the sets we have unstable regions bounded by $
2+2\rho\:>\:d\:>\:\frac{6+9\rho}{3+2^{-\rho}} $ (LR-SR) and $
2+2\beta+2\rho\:<\:d\:<\:2+2\beta+3\rho $ for $ \rho,\beta > 0 $
(LR-LR). Also both these fixed points give non-KPZ results:

\bea
z\mid_{{{U^*}^2}_{\rho,0}=0}\:=\:\frac{1}{3}\:(d+4-2\beta-
2\rho)
\eea
and
\bea
z\mid_{{{U^*}^2}_{\rho,\beta}=0}\:=\:2\:+\:\frac{(d-2-2\rho)
(d-2-3\rho)}{(3+2^{-\rho})d-6-9\rho}
\eea

and as such the phase diagram in $ \lambda_0,\:\lambda_{\rho},\:D_0,\:D_
{\beta} $ is actually on a 4-dimensional space with both axial and non-axial fixed points.
\par
Now use is made of self-consistent mode analysis (exact in the spherical
limit) to generate the non-KPZ exponents in this complex space where both the SR-LR noises are interacting with the SR-LR non-linearity.
\par
The Dyson's equation gives $ G^{-1}(\vec k,\omega) = -i \omega + \nu k^2 +
\Sigma(\vec k,\omega) $, where $ \Sigma(\vec k,\omega) $ is the self-energy
term and the corresponding scheme gives

\ber
\Sigma(\vec k,\omega)\:&=&\:\int\:\frac{d^d k^{\prime}}{{2\pi}^d}\:\frac{d\omega^
{\prime}}{2\pi}\:\vec k^{\prime}_{+}.\vec k^{\prime}_{-}\:v(2\vec
k^{\prime})v(\frac{3\vec k}{2}+\vec k^{\prime}) 
 \nonumber \\
& &G_0(\vec k^{\prime}_{-},\omega_{-}^{\prime})\:G_0(\vec
k^{\prime}_{+},\omega^{\prime})\: 
G_0(-\vec k^{\prime}_{+},-\omega_{+}^{\prime})\:2D(\vec
k^{\prime}_{+},\omega_{+}^{\prime}) 
\eer

The noise correlation function at the same order of accuracy $ (O(v^2)) $
gives

\ber
D(\vec k,\omega)\:&=&\:D_0(\vec k,\omega)\:+\:\mid {G_0}^2(\vec k,\omega)
\mid \:\int\:\frac{d^d k^{\prime}}{(2\pi)^d}\:\frac{d\omega^{\prime}}{2\pi}
\nonumber \\
& &{{k^{\prime}_{+}}^2}.{{k^{\prime}_{-}}^2}\:v^2(2\vec
k^{\prime})\:D({\vec k^{\prime}}_{+},{\omega}_{+}^ 
{\prime})\:D({\vec k^{\prime}}_{-},{\omega}_{-}^{\prime}) \nonumber \\
& &G_0(\vec k^{\prime}_{-},\omega_{-}^{\prime})\:G_0(-\vec
k^{\prime}_{-},-\omega_{-}^ 
{\prime})\:G_0(\vec k^{\prime}_{+},\omega_{+}^{\prime})\:G_0(-\vec
k^{\prime}_{+},-\omega_{+}^{\prime}) 
\eer

We now make the following scaling ansatz for $ \Sigma(\vec k,\omega) $ and
$ D(\vec k,\omega) $:

\bea
\Sigma(\vec k,\omega)\:=\:{\mid {\vec k}
  \mid}^z\:\psi(\frac{\omega}{{\mid {\vec k} \mid}^z}) 
\eea
\bea
D(\vec k,\omega)\:=\:{\mid {\vec k}
  \mid}^{-2\beta}\:\phi(\frac{\omega}{{\mid {\vec k} \mid}^z}) 
\eea

In eqn.(20) the noise strength $ D_0 $ will be renormalized only if the
second term is more singular than the first. If the second term in
eqn.(20) dominates in the long wavelength limit, eqn.(20) reduces to

\ber
D(\vec k, \omega) &\simeq& \mid G^2(\vec k,\omega) \mid \:
\int\:\frac{d^d k^{\prime}
}{{2\pi}^d}\:\frac{d\omega^{\prime}}
{2\pi}\:{{k^{\prime}_{+}}^2}.{{k^{\prime}_{-}}^2} \nonumber \\
& &v^2(2\vec k^{\prime})\:D({\vec
  k^{\prime}}_{+},{\omega}_{+}^{\prime})\:D({\vec k^{\prime}}_{-}, 
{\omega}_{-}^{\prime}) \nonumber \\
& &G(\vec k^{\prime}_{-},\omega_{-}^{\prime})\:G(-\vec
k^{\prime}_{-},-\omega_{-}^{\prime}) \:G(\vec k^{\prime}_{+},\omega_{+}^{\prime})\:G(-\vec
k^{\prime}_{+},-\omega_{+}^{\prime}) 
\eer

In the limit $ \omega \rightarrow 0 $, taking $ D(\vec k,0) = D_{\beta}\:
k^{-2\beta} $ and $ v(\vec k) = \lambda_{\rho}\:k^{-\rho} $, simple power
counting from eqns.(19) and (23) gives

\bea
z\:=\:\frac{1}{3}\:(d+4-2\beta-2\rho),\:\alpha\:=\:-
\rho\:+\:\frac{1}{3}\:(2-d+2\beta+2\rho).
\eea

This values of z and $ \alpha $ are seen to tally very well with our
dynamic RG derivations in the $ U_{0,0},\:
U_{\rho,0},\:U_{0,\beta},\:U_{\rho,\beta} $ space, where in either of
the two planes the non-KPZ values coincide with eqn.(24). However, the
$ U_{\rho,0},\:U_{\rho,\beta} $ plane is special in that it provides
two axial non-KPZ values only one of which appears in the
self-consistent results. The other value is actually an off-shoot of
the SM and SMB derivation [3] where the relevant LR non-linearity is interacting with the SR part of the noise. That way we are trodding
into a new domain of roughness where both the non-KPZ values of z (and
$ \alpha $) exist although no rough-rough phase transition apparently
takes place.
\par
In summary, we have started with a simple phenomenological equation
which incorporates a non-local term in its interaction spectrum and
while coupling with spatially correlated noise develops a set of
dynamic and growth exponents which contain a non-KPZ part. As and when
the SR part in the spectrum becomes unstable, these non-KPZ domains
surface up and completely different critical behaviour sets into
existence. With negative values for the long and short-ranged
non-linearities the phase diagrams are modified, with the LR roughness
now giving way to SR roughness without the appearance of any excluded
instability in the phases, the only exception being in the case of the
special plane already discussed. Also it would be worthwhile to mention that although the non-local contribution in the non-linearity never generates it's short-ranged counterpart, the non-local part in the noise spectrum develops a white noise.We reconfirm all these DRG
observations from a self-consistent technique and arrive at the same
set of non-KPZ exponents when the noise strength remains
non-renormalized. Finally comparisons with established results [3] and [4], which constitute only parts of our whole domain, provide expected results.

\section{Acknowledgement}
The author acknowledges partial financial support from C.S.I.R., India
and would sincerely like to express his gratitude to Prof. S. M.
Bhattacharjee and Prof. J. K. Bhattacharjee for many illuminating
discussions and suggestions. The names of Mr. Saugata Bhattacharyya, Mr. Partha Pratim Ray and Mr. Suman Banik also deserve special mention.

\section{Figure Captions}
\bigskip
Fig.1 $ \lambda_{\rho} $ vs $ \lambda_0 $ phase diagram. The solid lines along x-axis represent LR phases while the dotted line in (B) shows a smooth phase.\\
Fig.2 $ D_0 $ vs $ D_{\beta} $ flow In the $ U_{0,0} $ vs
$ U_{0,\beta} $ plane. X and Y are the axial and off-axial fixed points respectively.
\end{document}